# *One-Atom-Thick IR Metamaterials and Transformation Optics Using Graphene*

Ashkan Vakil and Nader Engheta*

University of Pennsylvania

Department of Electrical and Systems Engineering

Philadelphia, Pennsylvania 19104, U.S.A.

**Abstract**

Here we theoretically show, by designing and manipulating *spatially inhomogeneous*, *non-uniform* conductivity patterns across a single flake of graphene, that this single-atom-layered material can serve as a "one-atom-thick" platform for infrared metamaterials and transformation optical devices. It is known that by varying the chemical potential using gate electric and/or magnetic fields, the graphene conductivity in the THz and IR frequencies can be changed. This versatility provides the possibility that different "patches" on a single flake of graphene possess different conductivities, suggesting a mechanism to construct "single-atom-thick" IR metamaterials and transformation optical structures. Our computer simulation results pave the way for envisioning numerous IR photonic functions and metamaterial concepts—all on a "one-atom-thick" platform—of such we list a few here: edge waveguides, bent ribbon-like paths guiding light, photonic splitters and combiners, "one-atom-thick" IR scattering elements as building blocks for 'flatland' metamaterials, thin strips as flatland superlenses, and "one-atom-thick" subwavelength IR lenses as tools for Fourier and transformation optics.

* To whom correspondence should be addressed. E-mail: engheta@ee.upenn.edu

The fields of plasmonics, metal optics, metamaterials and transformation optics *(1-7)* have received considerable attention in recent years, since they offer a variety of schemes to manipulate electromagnetic fields into desired spatial patterns, suggesting exciting potential applications in various branches of engineering and applied science. Owing to their ability to support the surface-plasmon polariton (SPP) surface waves *(8)* in the infrared and visible regimes, the noble metals, such as silver and gold, have been popular constituent materials for constructing optical metamaterials *(1)*. From the macroscopic electromagnetic view point, the plasmonic characteristics are associated with the noble metals permittivity function exhibiting negative real part *(8)*. However, the difficulty in controlling and varying permittivity functions of the noble metals and the existence of material losses in them—especially at visible wavelengths—degrade the quality of the plasmon resonance and limit the relative propagation lengths of SPP waves along such metal-dielectric interfaces. These drawbacks, therefore, constrain the functionality of some of metamaterials and transformation optical devices. It is, therefore, laudable to search for new suitable materials for these purposes.

In recent years, there has been an explosion of interest in the field of graphene, which has exciting properties in electronic transport *(9-24)*. An important parameter in describing optical characteristics of graphene is the complex conductivity $\sigma_g = \sigma_{g,r} + i\sigma_{g,i}$, which depends on radian frequency $\omega$, charged particle scattering rate $\Gamma$ representing loss mechanism, temperature $T$, and chemical potential $\mu_c$. The chemical potential depends on the carrier density and can be controlled by a gate voltage, electric field, magnetic field and/or chemical doping *(11-20)*. One of the interesting properties of graphene is that the imaginary part of its conductivity, i.e., $\sigma_{g,i}$, can, under certain conditions, attain negative and positive values in different ranges of frequencies depending on the level of chemical potential *(13)*. Figures 1a and b depict the real and imaginary parts of the conductivity of a free-standing isolated graphene as a function of frequency and chemical potential, computed from the Kubo formula with $T = 3\,\text{K}$ and $\Gamma = 0.43\,\text{meV}$ 11, 13 ($\mu_c$ up to $400\,\text{meV}$ and the frequency range of 10 to 50 THz). As can be seen, there are regions of frequencies and chemical potentials (or gate voltages) for which $\sigma_{g,i} < 0$, while in other regions $\sigma_{g,i} > 0$.

What is the significance of $\sigma_{g,i}$ attaining positive or negative values in the context of metamaterials? To address this issue, we momentarily assume that graphene has a very small thickness $\Delta$ (Later we shall let $\Delta \to 0$). We point out that it is possible to associate an *equivalent* complex permittivity for this $\Delta$-thick graphene layer. Defining a *volume* conductivity for this $\Delta$-thick graphene layer as $\sigma_{g,v} \equiv \dfrac{\sigma_g}{\Delta}$, we can then write its *volume* current density as $\boldsymbol{J} = \sigma_{g,v}\boldsymbol{E}$. Assuming $\exp(-i\omega t)$ time harmonic variations, we can rearrange the Maxwell equation $\nabla \times \boldsymbol{H} = \boldsymbol{J} - i\omega\varepsilon_o\boldsymbol{E}$ for the $\Delta$-thick graphene layer as $\nabla \times \boldsymbol{H} = (\sigma_{g,v} - i\omega\varepsilon_o)\boldsymbol{E}$. Denoting the *equivalent* complex permittivity of the $\Delta$-thick graphene layer by $\varepsilon_{g,eq}$, we obtain $\varepsilon_{g,eq} \equiv -\dfrac{\sigma_{g,i}}{\omega\Delta} + \varepsilon_o + i\dfrac{\sigma_{g,r}}{\omega\Delta}$. We realize that for a one-atom-thick layer, one cannot define any bulk permittivity. However, here we have temporarily assumed that the thickness of layer is $\Delta$, associating an equivalent permittivity with this single layer. This approach allows us to treat the graphene sheet as a thin layer of material with $\varepsilon_{g,eq}$. At the end we let $\Delta \to 0$, and recover the one-atom-thick layer geometry. Specifically, we note that $\text{Re}(\varepsilon_{g,eq}) \equiv -\dfrac{\sigma_{g,i}}{\omega\Delta} + \varepsilon_o \approx -\dfrac{\sigma_{g,i}}{\omega\Delta}$ for a very small $\Delta$, and $\text{Im}(\varepsilon_{g,eq}) \equiv \frac{\sigma_{g,r}}{\omega\Delta}$. This interestingly shows that the real part of equivalent permittivity for this $\Delta$-thick graphene layer can be positive or negative depending on the sign of the imaginary part of the graphene conductivity. Therefore, when $\sigma_{g,i} > 0$, i.e., $\text{Re}(\varepsilon_{g,eq}) < 0$, a single free-standing layer of graphene does effectively behave as a very thin "metal" layer, capable of supporting a transverse-magnetic (TM) electromagnetic SPP surface wave. Other groups have also predicted the possibility of existence of SPP along the graphene *(13-16, 22)*; however, here we present a different method to highlight this feature.

It is known that a slab of a material with complex permittivity $\varepsilon_m$ with negative real part (e.g., Ag or Au) and with thickness $\Delta$, surrounded by free space can support an odd transverse-magnetic (TM) electromagnetic guided mode with wave number $\beta$ expressed as (25)

$$\coth\left(\sqrt{\beta^2 - \omega^2 \mu_o \varepsilon_m}\, \Delta/2\right) = -\frac{\varepsilon_m}{\varepsilon_o} \frac{\sqrt{\beta^2 - \omega^2 \mu_o \varepsilon_0}}{\sqrt{\beta^2 - \omega^2 \mu_o \varepsilon_m}}. \tag{1}$$

By substituting $\varepsilon_m$ with the equivalent permittivity of the $\Delta$-thick graphene layer derived above, and letting $\Delta \to 0$, we get $\beta^2 = k_0^2 \left[1 - \left(\frac{2}{\eta_0 \sigma_g}\right)^2\right]$, which is the dispersion relation of the TM SPP optical surface wave along a graphene layer obtained by several groups *(13-16)*. However, when $\sigma_{g,i} < 0$ (i.e., when $\text{Re}(\varepsilon_{g,eq}) > 0$) TM guided surface wave is no longer supported on the graphene *(13-16)*. Instead a weakly guided transverse-electric (TE) surface wave might be present (13-14). Figures 1c and 1d present also the complex wave number $\beta$ for such TM SPP, as a function of frequency and chemical potential for $T = 3\,\text{K}$ and $\Gamma = 0.43\,\text{meV}$. The quantity $\text{Re}(\beta)/\text{Im}(\beta)$, which is known as figure-of-merit, and the propagation length $1/\text{Im}(\beta)$ are also shown in Figs. 1e and 1f.

By comparing the graphene with a thin layer of noble metal such as Silver or Gold, one may count at least three major advantages for a graphene layer: 1) In the mid infrared (IR) wavelengths, the "loss tangent" in graphene at low temperatures can be much lower than in silver or gold at room temperature. For example, for a free-standing graphene at $T = 3\,\text{K}$, $\Gamma = 0.43\,\text{meV}$, and the chemical potential $\mu_c = 0.15\,\text{eV}$ for a signal with the frequency 30 THz, the ratio of $\sigma_{g,r}/|\sigma_{g,i}|$, which is equivalent of $\text{Im}(\varepsilon_{g,eq})/|\text{Re}(\varepsilon_{g,eq})|$, is about $1.22 \times 10^{-2}$. On the other hand, this parameter for silver *(26)* at 30 THz at room temperature is about $5.19 \times 10^{-1}$; 2) As can be seen in Fig. 1c, the real part of wave number $\beta$ for the TM SPP wave along the graphene is much larger than the wave number of the free space, $k_o \equiv \omega\sqrt{\mu_o \varepsilon_o}$. As a result, such a SPP wave is very tightly confined to the graphene layer, with guided wavelength $\lambda_{SPP}$ much shorter than free space wavelength $\lambda_o$, i.e., $\lambda_{SPP} << \lambda_o$, consistent with Ref *(16)*. For a layer of graphene with above characteristics, we have $\text{Re}(\beta) = 69.34 k_o$ and $\text{Im}(\beta) = 0.71 k_o$, resulting an impressive figure of merit of $\text{Re}(\beta)/\text{Im}(\beta) = 97.7$; 3) arguably, the most important advantage of graphene over thin metal layers is the ability to dynamically tune the conductivity of graphene by means of chemical doping or gate voltage, i.e., $E_{bias}$ in real time, locally and inhomogeneously. In other words, by using different values of $E_{bias}$ at different locations across the single graphene layer, in principle we can create certain desired conductivity pattern. We mentioned that at a given frequency a proper choice of chemical potential (or equivalently gate electric biasing field $E_{bias}$) can provide us with $\sigma_{g,i} > 0$ or $\sigma_{g,i} < 0$. Since the conductivity is directly related to the equivalent permittivity for the $\Delta$-thick graphene layer as described above, the conductivity variation results in variation of equivalent permittivity across the single sheet of graphene. This provides an exciting possibility for tailoring and manipulating infrared SPP waves across the graphene layer. Therefore the graphene can be considered as a single-atom-thick platform for manipulation of IR signals, providing a "flatland" paradigm for IR metamaterials and transformation optics. In the rest of this Letter, we present several scenarios in which the proper choice of IR conductivity spatial patterns across the graphene provides exciting novel possibilities for tailing, manipulating, and scattering IR guided wave signals on the graphene. These scenarios can be a starting point for having flatland metamaterials and "one-atom-thick" transformation optical devices with exciting functionalities.

To start, consider the numerical simulation of the SPP mode at 30 THz guided by a uniformly biased graphene layer in Fig. 2a. The TM SPP guided wavelength for this free-standing graphene is $\lambda_{SPP} = \lambda_o / 69.34 = 144.22$ nm . This highly compressed mode offers an effective SPP index of $n_{SPP} \equiv \beta_{SPP} / k_o = 69.34$ and has a relatively long propagation length $l_{prop} = 15.6 \lambda_{SPP} = 0.225 \lambda_o = 2.25 \mu m$. In addition to the uniformly biased scenario, we can also engineer the SPP to reflect and refract on the same sheet of graphene by varying the electric bias spatially. To achieve this goal, one may consider three possible methods: 1) A split gate structure to apply different bias voltages to different gates. This is shown schematically in Fig. 2b, in which the potentials $V_{b1}$ and $V_{b2}$, applied to two gate electrodes, are chosen to provide different chemical potential values of, e.g. $\mu_{c1} = 0.15$ eV and $\mu_{c2} = 0.065$ eV in the two halves of graphene. For sake of clarity, in Fig. 2b and also later in part of Fig.3 the "gate electrodes" are symbolically shown above the graphene at a small distance. In practice, however, the gates are usually located on the substrate beneath the graphene; 2) An *uneven* ground plane, i.e., by designing the specific profile of the ground plane underneath the dielectric spacer holding the graphene, one can achieve *nonuniform* static biasing electric field under the graphene while the voltage applied between entire sheet of graphene and the ground plane is kept fixed. This is schematically shown in Figs. 3b and 3d (also with more details in Fig S1 and S3). Since in this scenario the separation distance between the graphene and the ground plane is not uniform, the static electric field due to the single bias voltage between the graphene and ground plane is nonuniform, therefore the sheet of graphene experiences different local carrier densities and hence different chemical potentials at different segments; and 3) Inhomogeneous permittivity distribution near the top surface of the spacer holding a sheet of graphene; inhomogeneous distribution of permittivity generates a nonuniform static electric field under the graphene, creating inhomogeneous chemical potential and in turn inhomogeneous conductivity across the graphene (see Fig S2). It is worth to note that with current nanofabrication techniques, it seems that it is straightforward to achieve deeply subwavelength widths for the graphene region with a different conductivity value.

In Fig. 2b, with these bias arrangement the conductivity values of the two segments are, respectively, $\sigma_{g1} = 0.0009(31) + i0.0765(06)$ mS and $\sigma_{g2} = 0.0039(25) - i0.0324(30)$ mS. The "farther" half section with $\sigma_{g1,i} > 0$ supports a TM SPP, while the "closer" half with $\sigma_{g1,i} < 0$ does not. Therefore, if a TM SPP is launched in the farther-half section towards the interface of two sections, it reflects back at that "invisible" boundary line on the same graphene. Figure 2b shows the simulation results which do support this phenomenon. In this scenario the incoming and reflected SPPs are combined to form an oblique standing wave. The reflection of SPP at this line resembles the Fresnel reflection of a plane wave from a planar interface between two media. Here, however, such reflection occurs along an essentially "one-atom-thick" platform, with a little radiation loss due to the high confinement of SPP to the graphene. This case might also be analogous to the Fresnel reflection from a planar interface between a medium that supports a propagating wave (e.g., a medium with a real refractive index such as a dielectric) and another medium that does not support a propagating wave (e.g., a medium with no real index, such as a noble metal). Accordingly, on the graphene the Fresnel reflection results in a near complete reflection. The simulation results reveal an effective reflection at the boundary "invisible" line between the two segments.

Based on the analogy we just established, we can have a guided IR edge wave along the boundary line between these two sections. The numerical simulations show the presence of such guided IR edge wave, as shown in Fig. 3a. This special guided wave propagates along a "one-atom-radius" boundary line. By post processing the simulation results, we estimate the wavelength of the guided edge wave to be around $\lambda_{edge} = 61.5$ nm. This phenomenon might be relevant, and coupled to, the electronic behavior near the p-n junction edge on the graphene *(18-20)*.

By extending this idea, we propose a setting analogous to a conventional 3D metal-dielectric-metal waveguide, nonetheless on a "one-atom-thick" platform. Figure 3b presents simulation results for such a so-called 2D "metal-dielectric-metal" waveguide, on a "one-atom-thick" graphene. In this case there are three distinct regions on the graphene: two side regions with chemical potential $\mu_{c2}$ that have conductivity $\sigma_{g2} = 0.0039(25) - i0.0324(30)$ mS with $\sigma_{g2,i} < 0$, and a middle narrow "ribbon-like" section, with chemical potential $\mu_{c1}$ that has conductivity $\sigma_{g1} = 0.0009(31) + i0.0765(06)$ mS where $\sigma_{g1,i} > 0$. The uneven ground plane, schematically shown under the graphene in Fig. 3b, is a proposed method to achieve these two different chemical potentials—however, in the numerical simulation, the graphene is assumed to be free standing in vacuum, since the SPP is highly confined to the graphene—refer to Fig. S3 for a schematic of the idea of uneven ground plane. A guided SPP wave, bounded by the two boundary lines between the graphene sections, is present, as shown in Fig. 3b. The width of the ribbon-like path is 200 nm. Figure 3c shows a similar structure but with an arbitrarily chosen narrower width of 30 nm, showing that coupled edge wave exists in this ribbon-like "one-atom-thick" waveguide. This scenario may be realized by using an inhomogeneous distribution of permittivity of dielectric spacer (not shown here) under the graphene. This Figure shows that by spatially varying the bias arrangement, we can also bend the ribbon, and still maintain the bounded SPP guided through the bend. Additionally, Figure 3d demonstrates an IR splitter, which can be realized either by employing proper spatial distribution for the bias electric field—which can be achieved by proper design of uneven ground plane—or by applying different bias voltages to electrodes in a split gate scenario. The mechanism for guiding IR signals across the graphene offers exciting possibility for highly miniaturized, "one-atom-thick" photonic nanocircuitry with numerous potential applications in information processing at the nanoscale *(27)*.

Besides the mechanism described above, we can also have "one-atom-thick" scenarios analogous to 3D guided wave propagation in optical fibers with two different dielectric media of different refractive indices as the core and cladding. In other words, the two sections of graphene could be biased to support two TM SPP modes with two distinct effective indexes $n_{SPP,1}$ and $n_{SPP,2}$. We can then realize waveguiding effects similar to those illustrated earlier based on this notion (not shown here for the sake of brevity).

By exploiting the above concepts one can, as well, envisage 2D IR metamaterials and transformation optical devices on a single layer of carbon atoms. In Fig. 4a, we present our simulation results for the SPP propagation along a layer of graphene within which, an array of two-dimensional (2D) circular "patches" is created. These patches are biased at voltage $V_{b2}$ ($\sigma_{g2} = 0.0039(25) - i0.0324(30)$ mS with $\sigma_{g2,i} < 0$), while the rest of graphene is biased at $V_{b1}$ ($\sigma_{g1} = 0.0009(31) + i0.0765(06)$ mS with $\sigma_{g1,i} > 0$). We note that each circular patch acts a scatterer for the SPP surface wave, behaving as a "one-atom-thick" "flatland inclusion". The collection of these "inclusions" creates a 2D bulk flat metamaterials. The SPP interaction in a two-dimensional IR metamaterial is demonstrated in the simulation results shown in Fig. 4a. This numerical demonstration verifies that the proposed geometry may indeed be designed to be the 2D analog of the 3D metamaterials formed by collections of subwavelength metallic nanoparticles—metamaterials that can exhibit backward wave propagation *(28)*.

Furthermore, an example of an IR transformation optical device is presented in Fig. 4b. A "flat" version of a Luneburg lens is designed on the graphene by creating several concentric rings with specific conductivity values. These values may be obtained either by applying a proper set of bias voltages, by properly designing the uneven ground plane, or by fabricating inhomogeneous spacer permittivity. For instance, with a special configuration of bias arrangement, we can create, approximately, a graded conductivity pattern that provides the required effective SPP graded index for operation of the Luneburg lens. To find the corresponding conductivities for these concentric rings we use the discretised approximate expression $\sigma_{i,n} = \sigma_{i,out}\left[2 - (r_n + r_{n-1})^2/4\right]^{-1/2}$ where $\sigma_{i,n}$ and $r_n$ are, respectively, the imaginary part of conductivity and radius of the *n*th section and $\sigma_{i,out}$ is the imaginary part of conductivity of the "background" graphene on which the lens is created. As our simulation results in Fig. 4b reveal, the SPP generated from a "point-like" source is evolved into an approximately "one-atom-thick collimated beam" of SPP on the graphene, as analogously a conventional 3D Luneburg lens acts for a 3D beam. The diameter of the "flat" Luneburg lens in our simulation is about $1.5\ \mu m$, which is $0.15\lambda_o$—a notably subwavelength size. This example suggests that one can design various subwavelength IR devices (e.g., convex lens, concave lens, etc) on the graphene—a versatile platform for nanoscale Fourier optics *(27)* and other photonic signal processing methods.

Finally, a flat version of the "superlensing" effect *(29-30)* is shown in Fig. 4c. In this simulation, an IR source on a graphene, biased at voltage $V_{b1}$ (equivalently chemical potential $\mu_{c1}$), is situated near a "strip" region of graphene biased at $V_{b2}$ (or $\mu_{c2}$). With proper choice of biases (resulting in a required set of conductivity values for the strip and the region outside the strip) and proper adjustment of width of the strip and separation between the source and the strip, we can implement an approximate superlensing effect, as our simulation results verify this possibility.

In conclusion, our theoretical study of IR wave interaction with graphene suggests that the graphene can be used as a low-loss "one-atom-thick" platform for flatland IR metamaterials and transformative optics. The required inhomogeneous, nonuniform patterning of conductivity may be achieved by various techniques such as varying chemical potentials, creating uneven ground plane, fabricating inhomogeneous permittivity of spacer dielectric, applying gate electric or magnetic field, and/or utilizing heated atomic force microscopy technique to reduce oxide of a graphene oxide sheet *(21)*. This unique platform opens new vistas in nanoscale and microscale photonic information processing and photonic circuitry.

**Acknowledgements**

This work is supported in part by the US Air Force Office of Scientific Research (AFOSR) grant number FA9550-08-1-0220.

**Figures:**

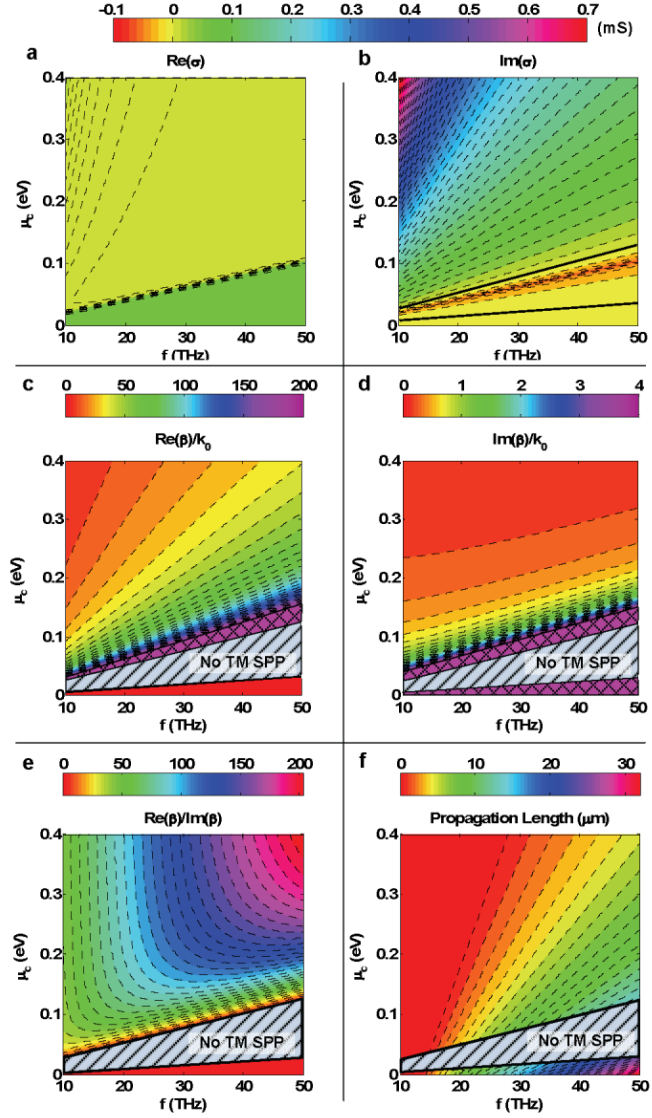

**Fig. 1**. Graphene conductivity and characteristics of the transverse-magnetic (TM) SPP surface waves. (**A**) Real part of the conductivity of a free-standing Graphene as a function of chemical potential $\mu_c$ and the frequency $f$, according to Kubo formula *(12, 14)* for $T = 3\,\mathrm{K}$, $\Gamma = 0.43\,\mathrm{meV}$. (**B**) Imaginary part of the conductivity for the same graphene sheet. **c,** Real part of the wave number $\beta$ of the TM SPP supported by the graphene with the same parameters mentioned in (**A**) and (**B**)—The purple cross-hatched area is corresponding to $\mathrm{Re}(\beta) \geq 200 k_0$.

(**D**) Imaginary part of the wave number $\beta$ of the TM SPP supported by the same graphene layer mentioned in (**A**) and (**B**). When $\text{Im}(\sigma_g) < 0$, no TM SPP is supported—The purple cross-hatched areas are corresponding to $\text{Im}(\beta) \geq 4k_0$. (**E**) The "figure-of-merit", defined as $\text{Re}(\beta)/\text{Im}(\beta)$, for the SPP mode as a function of $\mu_c$ and $f$. (**F**) The propagation length of the SPP mode, defined as $1/\text{Im}(\beta)$.

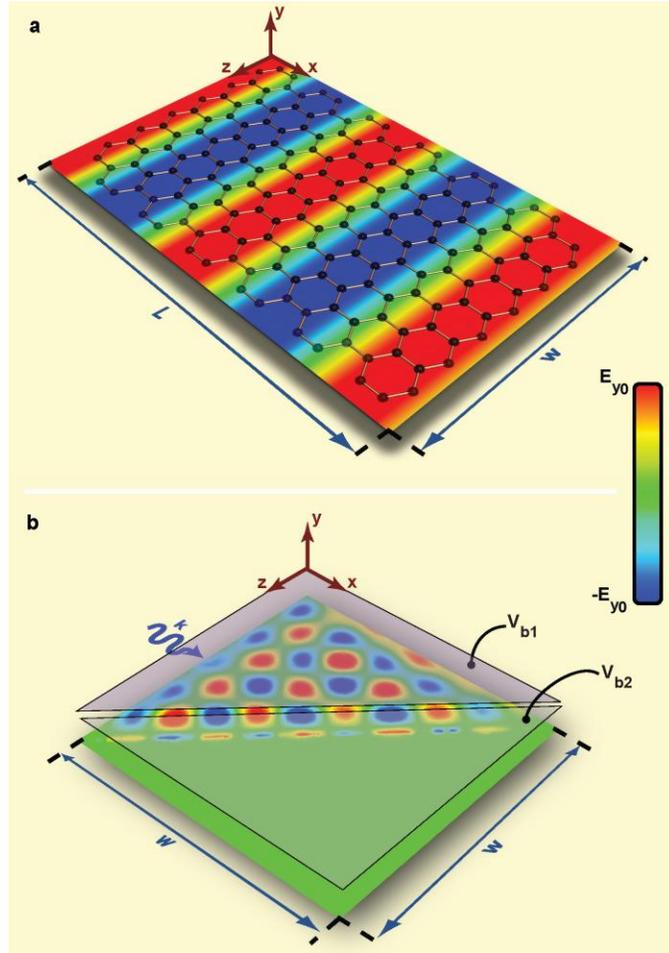

**Fig. 2**. (**A**) Simulation results showing the y-component of the electric field, $E_y$, for a TM SPP mode at 30 THz guided by a uniformly biased graphene layer with dimensions $L = 350$ nm, $w = 235$ nm ($T = 3$ K, $\Gamma = 0.43$ meV, $\mu_c = 0.15$ eV across the entire graphene). This chemical potential can be achieved, for example, by a bias voltage of 22.84 V across a 300-

nm $SiO_2$ spacer between the graphene and the Si substrate (but Si substrate and SiO2 spacer are not considered here). The SPP wavelength along the graphene, $\lambda_{SPP}$ is much smaller than free-space wavelength $\lambda_o$, i.e., $\lambda_{SPP} = 0.0144\lambda_o$. (**B**) Simulation results of the $E_y$ for the near total reflection of a TM SPP on a sheet of graphene with side $w = 800$ nm; two different gate bias voltages are applied. The $V_{b1}$ and $V_{b2}$ are chosen to provide different chemical potential values of $\mu_{c1} = 0.15$ eV and $\mu_{c2} = 0.065$ eV in the two halves of graphene (corresponding complex conductivity values are $\sigma_{g1} = 0.0009(31) + i0.0765(06)$ mS and $\sigma_{g2} = 0.0039(25) - i0.0324(30)$ mS). Section 1 supports a TM SPP, while Section 2 does not; see the text for complete explanation of this effect.

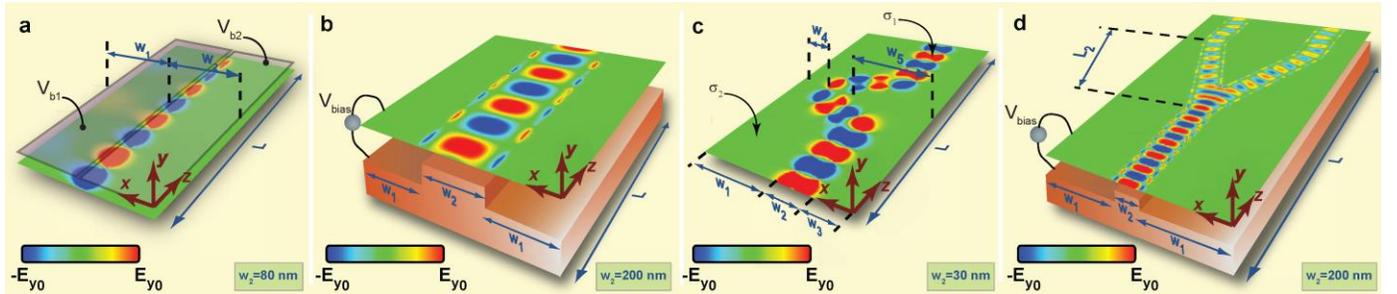

**Fig. 3.** (**A**) Distribution of $E_y$ (snap shot in time) of a guided IR edge wave at $f = 30\,THz$, supported along the boundary line between the two sections of the same sheet of graphene, which has two different conductivity regions, as in Fig. 2b ($L = 250$ nm and $w = 80$ nm). (**B**) Simulation results of $E_y$ (snap shot in time) for an IR guided wave at $f = 30\,THz$ along the ribbon-like section of graphene with the chemical potential $\mu_{c1}$, which may be achieved by heightening the ground plane underneath this region. The corresponding graphene conductivity of the ribbon is $\sigma_{g1} = 0.0009(31) + i0.0765(06)$ mS. This ribbon-like path is surrounded by the two other sections of graphene with chemical potential $\mu_{c2}$, which may be obtained by reducing the height of the ground plane beneath these regions. The corresponding graphene conductivity is $\sigma_{g2} = 0.0039(25) - 0.0324(30)$ mS ). The IR signal is clearly guided along this "one-atom-thick" ribbon. The computational region has the length $L = 560$ nm and total width $(W_1 + W_2 + W_3) = (120 + 120 + 120)$ nm. (**C**) Similar to panel b, but with the ribbon width of 30 nm, arbitrarily chosen to demonstrate the effects of coupling of two edge waves, providing another waveguiding phenomenon along this ribbon and also showing the bending effect. The computational region has the length $L = 370$ nm and total width $(W_1 + W_2 + W_3) = (120 + 30 + 60)$ nm while for the bent region $(W_4 + W_2 + W_5) = (30 + 30 + 120)$ nm. (**D**) Similar to panel b, but with the ribbon-like section split into two paths ($L_2 = 1077$ nm). The splitting of the SPP along this one-atom-thick structure is clearly seen in this simulation. The computational region shown has the length $L_1 = 2540$ nm and total width $(W_1 + W_2 + W_3) = (600 + 200 + 600)$ nm. Note different scale bars in different panels.

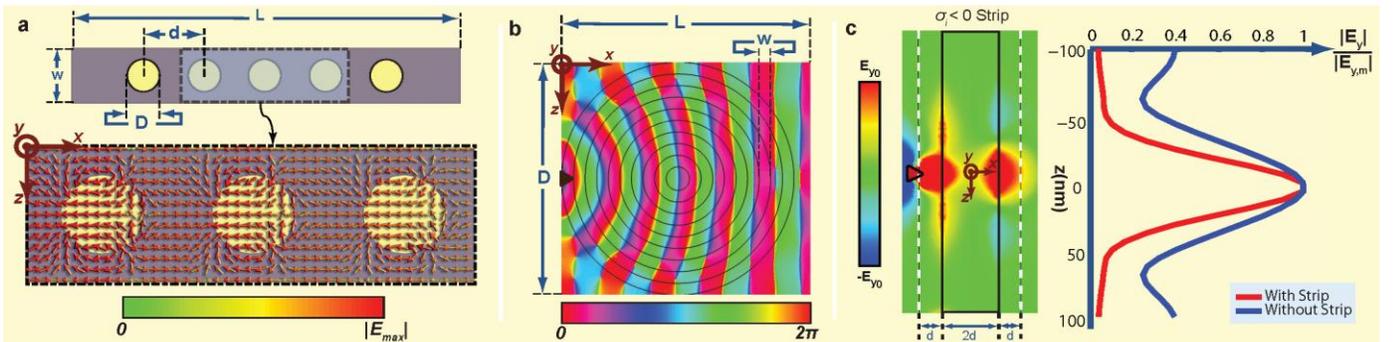

**Fig. 4** (**A**) Flatland Metamaterials: the snap shot in time of the electric field vectors for the SPP at $f = 30$ THz, shown on the x-z plane of the graphene, on which a 2D array of subwavelength circular patches is assumed. The patches conductivity is $\sigma_{g2}$, where $\text{Im}(\sigma_{g2}) < 0$. The rest of the graphene has the conductivity $\sigma_{g1}$ with $\text{Im}(\sigma_{g1}) > 0$. Only one row of the 2D periodic array is shown ($D = 30$ nm, $d = w = 55$ nm, $L = 370$ nm). Scattering of the SPP from these patches along the graphene leads to a bulk SPP property, yielding a flatland IR metamaterial. (**B**) Flatland Luneburg lens: Simulation results for the phase of $E_y$ of the SPP at $f = 30\,THz$ along the graphene, on which ten concentric annular regions with required conductivity values, according to Luneburg lens expression, are produced ($D = 1.5$ μm nm, $w = 75$ nm, $L = 1.6$ μm). Simulation results indicate such "one-atom-thick" Luneburg lens indeed collimates the SPP. (**C**) Flatland "superlens": Simulation results for $E_y$ of SPP at $f = 30\,THz$ on the graphene with a subwavelength strip region with conductivity $\sigma_{g2}$, while the rest of graphene has the conductivity $\sigma_{g1}$. The object—a point source—and image planes are assumed to be respectively 10 nm away from the left and right edges of the strip ($w = 2d = 20$ nm). The normalized intensity of $E_y$ at the image plane is shown for two cases with and without the strip (Normalization is with respect to their respective peak values). The subwavelength "focusing" is observable due to presence of the strip with conductivity $\sigma_{g2}$—$\text{Im}(\sigma_{g2}) < 0$.

**Supporting Material**

**Methods**

We have used commercially available full-wave electromagnetic simulator software, CST Studio Suite *(31)* in order to obtain the 3D numerical simulations presented in Figs 2, 3 and 4. For the purpose of our 3D simulation, the thickness of graphene is assumed to be 1 nm, although other extremely small values for this thickness lead to similar results. (We have assumed 1 nm thickness for graphene and assigned the corresponding permittivity in our simulations. Note that as long as the chosen thickness is extremely small compared to the wavelength, this particular choice is not essential—we could assume thickness of 0.5 nm and find the corresponding permittivity value.) Due to the large difference in the dimensions of the graphene layer (thickness vs. width and length), and also due to the special form of the conductivity function of graphene, we have chosen frequency-domain Finite Element Method (FEM) solver of CST. This solver solves the problem for a single frequency at a time. For each frequency sample, the linear equation system is solved by an iterative solver. Adaptive tetrahedral meshing with a minimum feature resolution of 0.5 nm has been used in the simulations. A point source (equivalent of an infinitesimal dipole antenna) has been utilized as the excitation of the structures. All the simulations reached proper convergence; a residual energy of $10^{-5}$ of its peak value has been obtained in the computation region. In order to absorb all the energy and to have approximately zero-reflection boundary, on the receiving side, in all the simulations a technique similar to the well-known Salisbury Shield method has been implemented (with proper modifications for a TM SPP mode). Depending on the nature of the problem, perfect magnetic conducting (PMC), perfect electric conducting (PEC) and open boundary conditions have been applied to different boundaries, to mimic the two-dimensionality of the geometry.

**Figures**

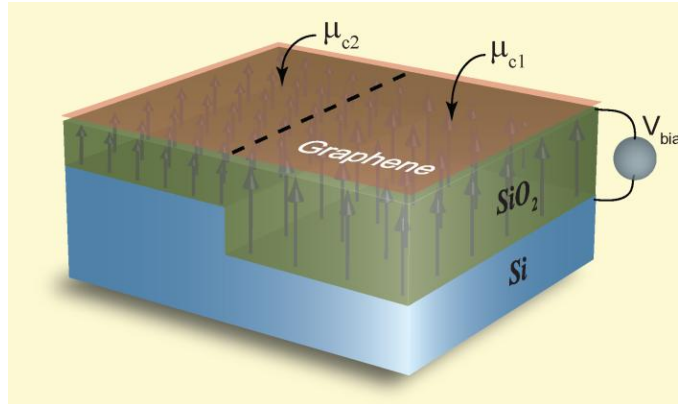

Fig. S1. Schematic of the idea of *uneven* ground plane underneath the graphene layer in order to construct inhomogeneous conductivity pattern across a single flake of graphene. By biasing the graphene with a single static voltage, the static electric field is distributed according to the height of the spacer between the graphene and the uneven ground plane, leading to the unequal static electric field. This results in unequal carrier densities and chemical potentials $\mu_{c1}$ and $\mu_{c2}$ on the surface of the single graphene and thus different conductivity distributions across the graphene.

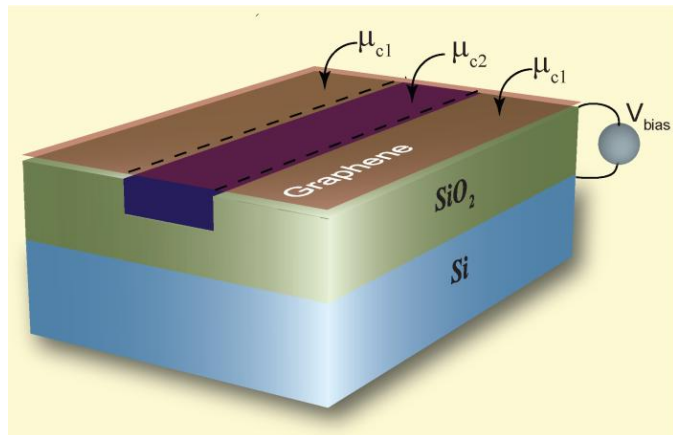

Fig. S2. Sketch of another idea to create inhomogeneous conductivity across a single sheet of graphene. Here several dielectric spacers with unequal permittivity functions are assumed to be used underneath of the graphene. This can lead to unequal static electric field distributions, resulting in inhomogeneous conductivity patterns across a single sheet of graphene.

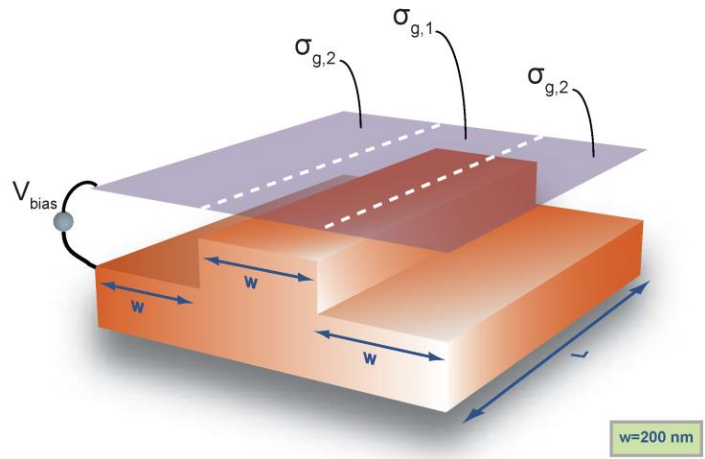

Fig. S3. Another schematic of the uneven ground plane idea to create inhomogeneous conductivity across a single sheet of graphene. This idea is also sketched in our Fig. 3.